\documentclass[9pt,conference]{IEEEtran}

\usepackage[preprint]{waspaa25}

\usepackage{bm} 
\usepackage{multirow}
\usepackage{comment}

\title{Source Separation of Small Classical Ensembles: Challenges and Opportunities}

\name{
 \parbox{\linewidth}{\centering
    Gerardo Roa-Dabike$^{4,1}$~
    Trevor J. Cox$^{1}$~
    Jon P. Barker$^{4}$~
    Michael A. Akeroyd$^{2}$~
    Scott Bannister$^{3}$ \\
    Bruno Fazenda$^{1}$~
    Jennifer Firth$^{2}$~
    Simone Graetzer$^{1}$~ 
    Alinka Greasley$^{3}$~
    Rebecca R. Vos$^{1}$~
    William M. Whitmer$^{2}$
    }
}
\address{
    $^{1}$University of Salford\quad
    $^{2}$University of Nottingham\quad
    $^{3}$University of Leeds\quad
    $^{4}$University of Sheffield
}

\begin{document}

\maketitle

\begin{abstract}


Musical source separation (MSS) of western popular music using non-causal deep learning can be very effective. In contrast, MSS for classical music is an unsolved problem. Classical ensembles are harder to separate than popular music because of issues such as the inherent greater variation in the music; the sparsity of recordings with ground truth for supervised training; and greater ambiguity between instruments. The Cadenza project has been exploring MSS for classical music. This is being done so music can be remixed to improve listening experiences for people with hearing loss. To enable the work, a new database of synthesized woodwind ensembles was created to overcome instrumental imbalances in the EnsembleSet. For the MSS, a set of ConvTasNet models was used with each model being trained to extract a string or woodwind instrument. ConvTasNet was chosen because it enabled both causal and non-causal approaches to be tested. Non-causal approaches have dominated MSS work and are useful for recorded music, but for live music or processing on hearing aids, causal signal processing is needed. The MSS performance was evaluated on the two small datasets (Bach10 and URMP) of real instrument recordings where the ground-truth is available. The performances of the causal and non-causal systems were similar. Comparing the average Signal-to-Distortion (SDR) of the synthesized validation set (6.2 dB causal; 6.9 non-causal), to the real recorded evaluation set (0.3 dB causal, 0.4 dB non-causal), shows that mismatch between synthesized and recorded data is a problem. Future work needs to either gather more real recordings that can be used for training, or to improve the realism and diversity of the synthesized recordings to reduce the mismatch. Furthermore, while the SDRs for the synthesized validation set are promising, there is still scope to further improve the models or to expand the training sets to improve performance.

\end{abstract}
\section{Introduction}
\label{sec:intro}

People with hearing loss can struggle to identify and distinguish instruments in classical ensemble music \cite{hake2023development, moore2016effects, siedenburg2020can}. Could music source separation (MSS) help? If MSS can separate music recordings into individual instrument lines, a remix can be made that increases the salience of some instruments. For instance, amplification, EQ or dynamic range compression could be applied to some instruments before remixing.

The Cadenza Project is running machine learning challenges to improve music processing for people with hearing loss. This is needed because 68\% of hearing aid users report difficulties when listening to music \cite{Greasley2020}. The second Cadenza Challenge (CAD2) included a task to rebalance musical instruments in a small classical ensemble \cite{roadabike2024}.

MSS is an established area of research and there have been previous demixing challenges \cite{sisec2018, vaisberg2019qualitative, Fabbro-2024}. These have focused on non-causal separation of pre-recorded, studio-produced Western popular music. High performance can be achieved with the Signal-to-Distortion-Ratio (SDR) exceeding 9 dB \cite{10446651, 10446843, 10096956} for the MUSDB18 benchmark \cite{musdb18-hq}.

People with hearing loss tend to be older and listen to more classic music \cite{bonneville2013music}. If MSS is to help with music processing in hearing aids, it needs to work with classical music. But classical music offers a number of additional challenges for MSS compared to Western popular music. These include: (i) A lack of datasets to train deep learning algorithms. (ii) Classical ensembles often include more instruments. (iii) Some instruments in an ensemble have similar timbres that are harder to distinguish (e.g. violin and viola). (iv) Classical ensemble recording is usually done with all musicians in a performance space, which adds room acoustics to the signals.

MSS for classical music is much less explored than popular music. Chiu et al. \cite{9287146} trained an Open-Unmix model \cite{stoter19} to separate piano-violin duets. 
For 6 tracks from the MedleyDB dataset, they achieved SDRs of 9.66 and 1.56 dB for piano and violin, respectively. Sarkar et al. created the synthesized EnsembleSet data and then used it to train a duet separator using a dual-path transformer architecture \cite{sarkar2022ensembleset}. For evaluation, they extracted duets of real recordings from the URMP dataset \cite{li2018creating} and got an SDR of 11.37 dB. 
A recent paper created SynthSOD, a synthesized dataset that is better balanced across instruments than Ensembleset \cite{10839019}. To test their new dataset, Garcia-Martinez et al. trained four independent Open-Unmix separator models for (i) strings, (ii) woodwind, (iii) brass, and (iv) percussion. Tests on the URMP data gave SDRs ranging from 0.0 dB for the tuba to 4.51 dB for the cello. These lower SDRs show that MSS with more than two instruments and real recorded test sets is still unsolved for classical music.

Finally, previous MSS work has mostly developed non-causal algorithms. This is fine for recorded music, but for live music and hearing aids, causal approaches are needed.

This paper arises from the MSS part of the CAD2 Cadenza challenge for classical music ensembles. (The remixing and hearing aid amplification parts of the challenge are not reported.) Below, the data are outlined, including the new synthetic woodwind dataset developed for the challenge, and the causal and non-causal MSS models used that formed the challenge baseline. Results follow, along with discussions around the unsolved problems in MSS for classical ensembles.

\section{Materials and Methods}
\label{sec:materials_and_methods}

\subsection{Datasets}

Real recordings of classical ensembles with completely isolated instrument tracks are scarce, see \cite{10839019} for a summary. They are needed so the ground truth of the separated instrument signals are known. The recorded datasets that do exist are too small for training a deep neural network (DNN). Consequently, systems were trained on synthesized audio and evaluated on real recordings.

For strings, we used the synthesized EnsembleSet \cite{sarkar2022ensembleset} during training. This contains 80 pieces from 17 composers. EnsembleSet has synthesized audio for 11 different instruments and 18 virtual microphones in a concert hall. We used the professionally mixed Mix\_1 render, which is a balance of the Decca tree, outriggers, ambient, balcony, mids, and close microphones. One limitation is that there are relatively few woodwind examples, and so the set has an unbalanced representation of the different instruments. Hence, we complemented with the CadenzaWoodwind, a new dataset of synthesized woodwind quartets \cite{CadenzaWoodwind}. This audio was generated by selecting 19 music scores from the OpenScore String Quartet corpus \cite{gotham2023openscore}. These were rendered for two ensembles of (i) flute, oboe, clarinet and bassoon; and (ii) flute, oboe, alto saxophone and bassoon. Synthesis was done by a professional music producer using virtual instruments that interpreted the expression markings in the score. Convolution reverberation simulated a performance space. 

For the validation set, a subset of tracks from the training dataset was held out. This contained 8 tracks from EnsembleSet and 6 from CadenzaWoodwind. These tracks were divided into consecutive 15-second segments and split to generate 261 samples, totalling just over one hour of data. Validation samples included 87 samples of string quartets from EnsembleSet and woodwind quartets from CadenzaWoodwind, and random mixtures (174 samples) combining string and woodwind instruments because mixed ensembles appear in the evaluation data. This should also aid training through increased diversity \cite{jeon2024does}.

\begin{figure}
    \centering
    \includegraphics[width=1\linewidth]{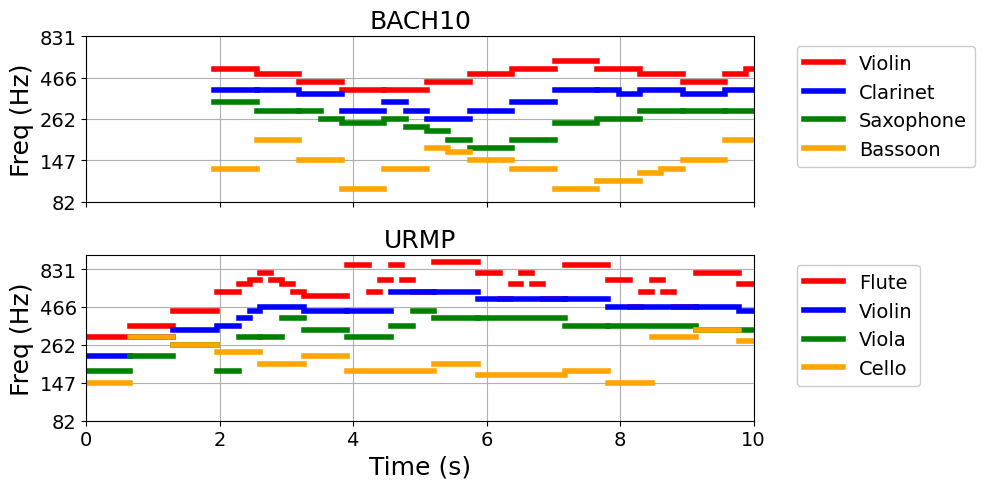}
    \caption{First 10 seconds of a quartet from Bach10 and URMP datasets.}
    \label{fig:eval_sets}
\end{figure}

For evaluation, we used the Bach10 \cite{duan2011soundprism} and URMP \cite{li2018creating} datasets. Bach10 has 10 four-part J.S.~Bach chorales (bass, tenor, alto and soprano), performed on bassoon, alto saxophone, clarinet, and violin. Each instrument was recorded in isolation in an anechoic chamber. To maintain synchronization, the performers listened to recordings of the other parts through headphones while playing. The URMP dataset has a total of 44 duets, trios, quartets and quintets. The pieces span works from 19 composers, including Mozart, Tchaikovsky and Beethoven, and are performed using a combination of 14 different instruments. Due to their low representation in the data, we excluded pieces featuring double bass and brass. As with Bach10, for each ensemble, instruments were recorded individually in an anechoic chamber. To maintain instruments' synchronization throughout the entire piece, performers were presented with a video recording of a conductor and an audio recording of a piano track through headphones. While immensely valuable, both datasets suffer from the musicians playing separately, for example, at times the instruments are not in tune with each other.

Both datasets have mono recordings of isolated instruments in anechoic conditions. To simulate a real listening scenario, we have taken these and created stereo versions in small halls using convolution reverb based on Ambisonic impulse responses from the OpenAIR database \cite{shelley2010openair}. The instruments were spaced at 10\textdegree~ apart, achieved by a rotation in the B-format representation. The B-format impulse response was then converted to mid-side stereo \cite{zotter2019ambisonics} before convolution. To examine the effects of reverberation, we also created stereo anechoic versions by applying gains based on the mid-side stereo formulations. 

Figure \ref{fig:eval_sets} shows a MIDI representation of the first 10 seconds of two quartets from the Bach10 and URMP datasets. The frequency ranges of the instruments overlap, and this characteristic is also preserved in the synthesized data.

We combined all lines of the same instrument within each ensemble. This is because the models were trained to separate a target instrument from an ensemble, rather than distinguish between multiple lines played by the same instrument (e.g. violin\_1 and violin\_2 in a string quartet). Since these instruments occupy the same frequency band and have the same timbre, they can only be separated with information from the score or with a DNN that has a more sophisticated understanding of polyphonic music. 

Table \ref{tab:rebalancing_dataset} shows the list of the eight 
instruments in our study
and the number of songs and duration in minutes for both the evaluation and training sets. The evaluation of each MSS model was performed using only the tracks that contain the target instrument in the datasets. As a result, the number of signals used to compute the SDRs varies between instruments. 

The SDR for each stereo sample was computed using the museval library \cite{sisec2018}. We first calculated the SDR over non-overlapping 1-second frames for each channel, then took the median across channels for each frame. Then, we computed the median across all frames within a track. The reported SDR corresponds to the median across all tracks.

\begin{table}[!htb]
    \centering
    \caption{Summary of the instruments in the training and evaluation datasets indicating the number of tracks containing the instruments and the total duration in parenthesis. `Training' includes the training and validation sets.}

        \begin{tabular}[t]{lcccc}
        \toprule
        \multirow{3}{*}{\textbf{Instrument}} & \multicolumn{2}{c}{\textbf{Evaluation}} & \multicolumn{2}{c}{\textbf{Training}}  \\
         \cmidrule(lr){2-3}\cmidrule(lr){4-5}
                & \multirow{2}{*}{\textbf{Bach10}} & \multirow{2}{*}{\textbf{URMP}} &  \multirow{2}{*}{\textbf{EnsembleSet}} & \textbf{Cadenza}   
                \\
                &&&& \textbf{Woodwind}\\
        \midrule
        Bassoon  & 10 (6 min) & 2 (1 min) & 2 (6 m) & 19 (536 min) \\
        Cello  & - & 7 (13 min) & 77 (369 min) & - \\
        Clarinet  & 10 (6 min) & 6 (10 min) & 2 (14 min) & 19 (536 min) \\
        Flute  & - & 11 (16 min) & 1 (6 min) & 19 (536 min) \\
        Oboe  & - & 3 (3 min) & 4 (33 min) &  19 (536 min) \\
        Saxophone  & 10 (6 min) & 5 (6 min) & - & 19 (536 min) \\
        Viola  & - & 9 (17 min) & 73 (358 min) & - \\
        Violin  & 10 (6 min) & 20 (31 min) & 79 (703 min) & -\\
        \bottomrule
        \end{tabular}
        
    \label{tab:rebalancing_dataset}
\end{table}
\subsection{MSS model}
\label{sec:experiment}

\begin{figure}[!t]
    \centering
    \includegraphics[width=1\linewidth]{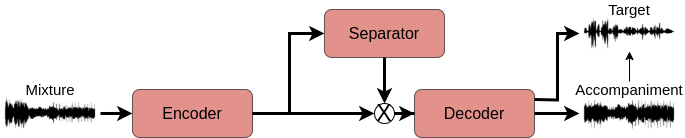}
    \caption{Block diagram of the TasNet architecture.}
    \label{fig:TasNet_arquitecture}
\end{figure}

The MSS was done by eight DNN models. Each was trained to extract one of the instruments shown in Table \ref{tab:rebalancing_dataset} alongside the rest of the music. These were ConvTasNet models \cite{convtasnet}. We selected this model because it can be easily switched between causal and non-causal modes. ConvTasNet is a fully convolutional source separation architecture composed of a convolutional encoder, a separator module composed of stacked 1-D convolutional blocks, and a transposed convolutional decoder (Figure \ref{fig:TasNet_arquitecture}). This architecture was originally designed for speaker separation from a mono-audio mixture. The version we used had already been adapted to popular music. All models were trained using the same configuration: stereo signals, 8 convolutional blocks (X in ConvTasNet) per repeat, and 3 repeats (R in ConvTasNet).

The models were trained for 200 epochs with 4 samples per batch. Early stopping was implemented after 20 epochs if there was no improvement in validation loss. We used the Adam optimizer and an F1 loss function. The initial learning rate was set at $1 \times 10^{-3}$ and was reduced by half after 5 epochs when the validation loss was not showing improvement. Audio samples were 3 seconds long and randomly selected within each audio track. For data augmentation, we applied random gain per stem and channel swapping.

\begin{table*}[!tb]
    \centering
    \caption{The SDRs (dB) for the different separated instruments for non-causal and causal models. Validation and evaluation (Bach10, URMP) results shown. 'Anech': stereo data with no reverberation. `Reverb': stereo data with reverberation. `Synth': synthesised samples. `Ref': the SMR for the reverb signals.}

        \begin{tabular}[t]{lccccccccccccc}
        \toprule
        
         & \multicolumn{5}{c}{\textbf{Bach10}} & \multicolumn{5}{c}{\textbf{URMP}} & \multicolumn{3}{c}{\textbf{Validation}}\\
         \cmidrule(lr){2-6}\cmidrule(lr){7-11}\cmidrule{12-14} 
         && \multicolumn{2}{c}{\textbf{Causal}} & \multicolumn{2}{c}{\textbf{Non-causal}} && \multicolumn{2}{c}{\textbf{Causal}} & \multicolumn{2}{c}{\textbf{Non-causal}} && \textbf{Causal} & \textbf{Non-causal} \\
         \cmidrule(lr){3-4}\cmidrule(lr){5-6}\cmidrule(lr){8-9}\cmidrule(lr){10-11} \cmidrule(lr){13-13}\cmidrule(lr){14-14}
         & \textbf{Ref} & \textbf{Anech} & \textbf{Reverb} & \textbf{Anech} & \textbf{Reverb} & \textbf{Ref} & \textbf{Anech} & \textbf{Reverb} & \textbf{Anech} & \textbf{Reverb} & \textbf{Ref} & \textbf{Synth} & \textbf{Synth} \\
   
        \midrule
Bassoon  & -5.338 & 0.263 & 0.142 & 0.235 & 0.075 & -6.727  & 1.222  & 0.299  & -0.055 & -0.845 & -18.292 & 0.815  & 1.710  \\
Cello    & -      & -     & -     & -     & -     & -3.458  & 0.191  & 0.268  & 0.077  & 0.379  & 0.654   & 12.580 & 13.171 \\
Clarinet & -5.047 & 0.109 & 0.122 & 0.661 & 0.022 & -3.168  & -0.085 & 0.658  & 0.896  & 1.348  & -9.353  & 3.227  & 3.996  \\
Flute    & -      & -     & -     & -     & -     & -5.547  & 0.217  & 0.557  & 0.669  & 1.943  & 1.358   & 7.456  & 7.911  \\
Oboe     & -      & -     & -     & -     & -     & -12.294 & 0.000  & 0.000  & 0.000  & 0.000  & -8.391  & 6.633  & 6.087  \\
Alto sax & -4.743 & 0.000 & 0.000 & 0.000 & 0.000 & -3.270  & 0.000  & 0.000  & 0.000  & 0.000  & -10.928 & 5.200  & 5.789  \\
Viola    & -      & -     & -     & -     & -     & -7.112  & 0.199  & 0.250  & 0.384  & 0.398  & -12.839 & 5.073  & 5.313  \\
Violin   & -5.780 & 1.061 & 0.666 & 1.440 & 0.385 & -7.485  & 0.550  & 0.346  & 0.178  & 0.020  & -2.389  & 8.611  & 11.388 \\
\cmidrule{2-14}
Average  & -5.227 & 0.358 & 0.233 & 0.584 & 0.121 & -6.133  & 0.287  & 0.297  & 0.269  & 0.405  & -7.523  & 6.199  & 6.921 \\
        
        \bottomrule
        \end{tabular}
        
    \label{tab:separation_results}
\end{table*}

\section{Results and Analysis}
\label{sec:results}

Table \ref{tab:separation_results} presents the SDR scores in dB for each instrument. For each evaluation dataset, we also report the SDR for an anechoic ('Anech') stereo rendition of the tracks to assess the influence of reverberation, by comparison with the signals containing small hall reverberation (`Reverb'). We also include a reference (`Ref') signal-to-music ratio (SMR), defined as the SDR computation with the target instrument as the signal and the full mix of all instruments as the distortion/music. This serves as a measure of how prominent the original instrument was in the original mix.

\subsection{Validation}

For the synthesized validation data, some separation has been achieved, as indicated by the average SDR being 14.1 dB higher than the reference SMR. The performance is comparable to previous work using an X-UMX MSS model \cite{sawata2021all} and the SynthSOD synthesized data for training and evaluation for ensembles of up to 5 instruments \cite{10839019}.

Comparing causal and non-causal models, the SDR averaged across instruments was 6.1 and 6.9 dB, respectively. The difference is not statistically significant (t=1.53, df=1756, p=0.13). It might be expected that the causal models would perform worse than the non-causal models, because non-causal processing has access to both past and future samples, whereas causal processing can only use previous and current audio to predict future samples. However, that was not the case. Furthermore, given that causal models are more useful because they can be applied to live music in real time, future work might be best focussed on these.

\begin{figure*}[htb] 
    \centering
    \begin{minipage}[b]{0.49\textwidth}
        \centering
        \includegraphics[width=\linewidth]{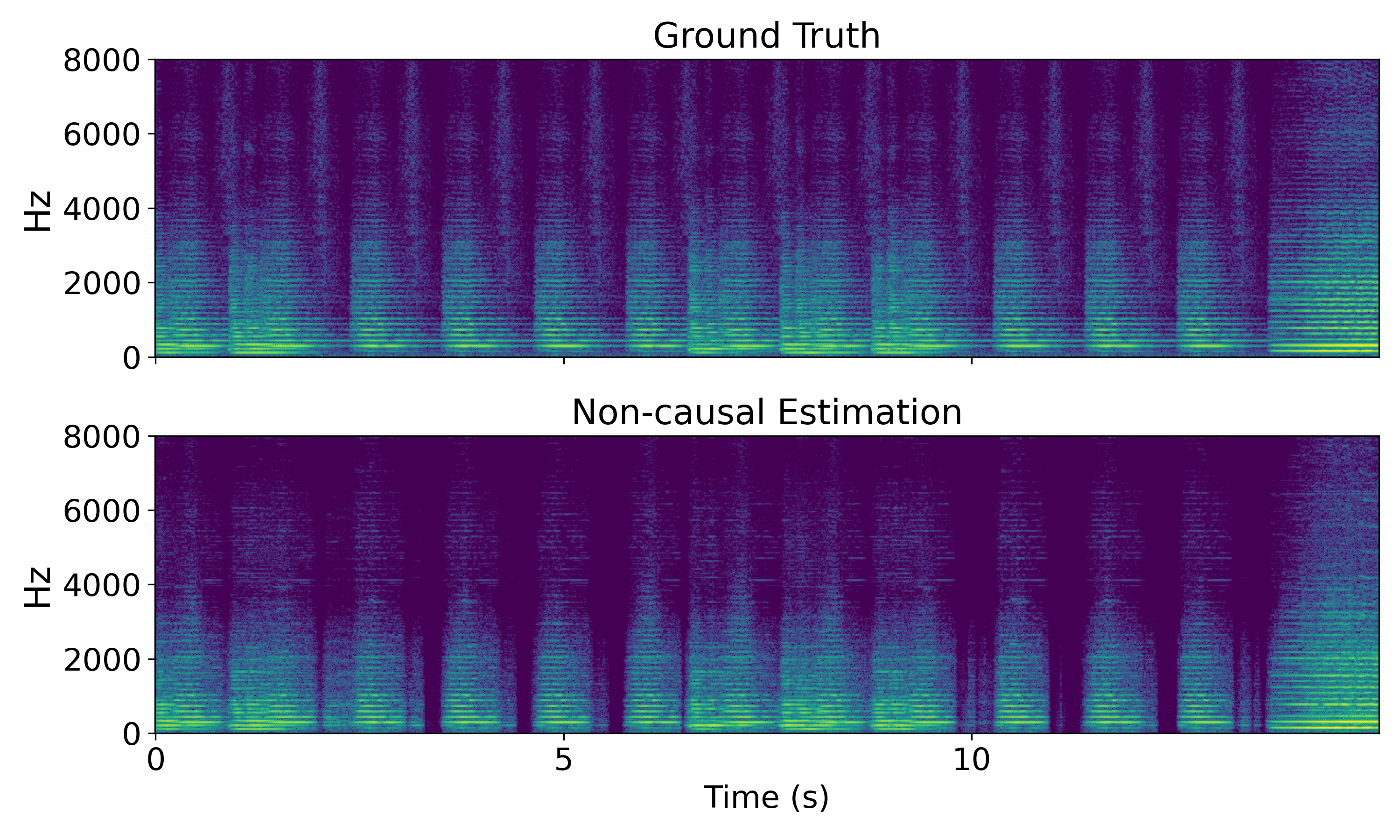}
        \caption{Ground Truth and estimation of cello from one validation sample.}
        \label{fig:cello}
    \end{minipage}
    \hspace{0.1cm} 
    \begin{minipage}[b]{0.49\textwidth}
        \centering
        \includegraphics[width=\linewidth]{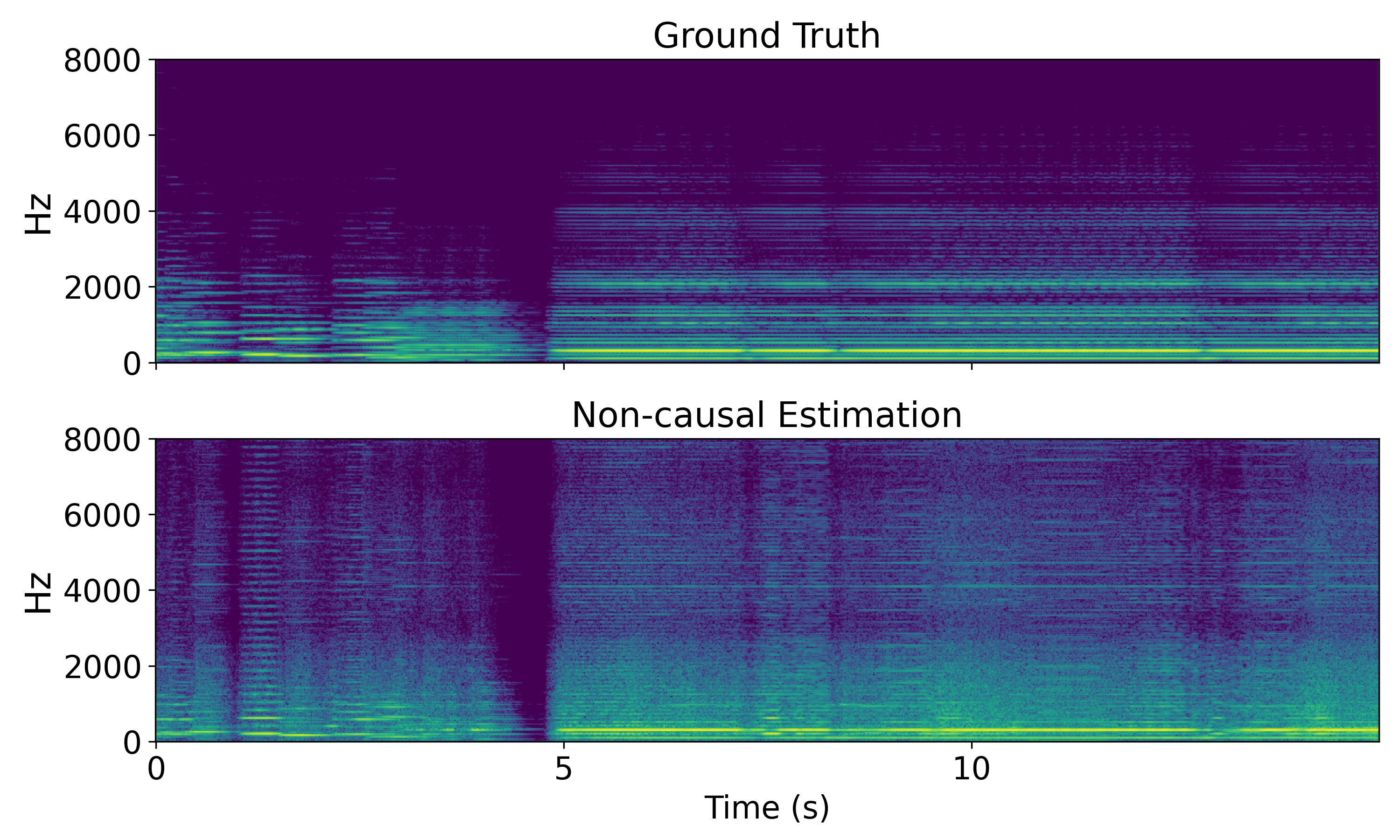}
        \caption{Ground Truth and estimation of bassoon from one validation sample.}
        \label{fig:bassoon}
    \end{minipage}%
      
\end{figure*}

The separation performance varied greatly across instruments, with the highest SDR being 13.1 dB for the cello (non-causal model) and the lowest 0.8 dB for the bassoon (causal model). 

For the cello, the models achieved good separation but failed to separate some of the reverberation occurring at the end of the notes (e.g. Figure \ref{fig:cello}), abruptly ending the notes.

Informal listening indicated that for the bassoon many notes were missing and often misclassified as clarinet or saxophone. For those that were correctly classified, most of the audio recovered was below 2000 Hz. In contrast, above 2000 Hz, the estimation introduced interference from other sources (e.g. Figure \ref{fig:bassoon}). There is no correlation between the frequency range of the instruments and their separation performance. There is also no significant correlation between the average SDR for each instrument and the reference SMR. The string instruments had an average SDR of 9.0 dB across the causal and non-causal systems, whereas the average for the woodwinds was significantly smaller at 4.5 dB (t=7.81; df=1756, p$<$0.001). This shows that the string instruments from EnsembleSet are easier to separate than the woodwinds from CadenzaWoodwind. This might be due to the different timbres of the instruments. Lending weight to this idea, a similar difference was found with the synthesized SynthSOD dataset, with an average SDR across violin, viola and cello of 7.1 dB SDR, and across flute, clarinet, oboe and bassoon 3.4 dB (evaluation on SynthSOD ensembles up to 5 instruments, training on SynthSOD) \cite{10839019}. But, the contrasting performance between strings and woodwinds in our study could also be due to differences in the virtual instruments and reverberation simulation used to create the two datasets.

\subsection{Evaluation}

The performance on the real recording with added hall reverberation from the Bach10 and URMP datasets was much poorer than that of the synthesized validation set, with average SDRs across instruments ranging from 0.12 to 0.58 dB. Although some separation has occurred, the MSS system is not working well enough, as the signal is roughly at the same level as the distortion. Based on listening to a few samples, it seems that the most common error was the misclassification of notes from similar instruments. Notable exceptions are the oboe and saxophone, which consistently obtained an SDR of only 0 dB across causality and acoustic conditions. For the saxophone, inspection of the separated signals revealed that estimations were almost silent for causal, non-causal, anechoic and reverberant conditions. This suggests a poor ability of the models to separate the saxophone effectively. Maybe the synthesized saxophone did not accurately capture the characteristics of the real instrument. In contrast, for the oboe, while the anechoic condition also led to nearly silent outputs, the reverberant condition produced distorted signals that roughly matched the ground truth, indicating some degree of separation.

For the Bach10 dataset with reverberation, all instruments have a similar original SMR with an average of -5.2 dB, indicating that the average masking for each instrument is comparable. The separation performance of each instrument varies, however. The best result was for the violin, with a over 6 dB improvement over the SMR. For the reverberated URMP dataset, the reference SMR varies across instruments, indicating different average masking -- with the oboe being the most masked, and the clarinet the least masked. As in the validation set, the flute achieved the highest performance among the woodwinds. Due to the small number of samples, results for bassoon and oboe should be interpreted with caution.

Garcia-Martinez et al. \cite{10839019} also used the URMP dataset for evaluation. They achieved an SDR of 1.01 dB across the 8 instruments that are common with our study. The average SDR of our system for the same instruments is slightly worse at 0.3 and 0.4 dB for the causal and non-causal systems.

The average SDR across instruments, models and tracks was 0.48 dB for the anechoic case and 0.44 dB for the reverberated signals. It might be expected that reverberation would make instruments harder to separate because it blurs the temporal and spectral features, but this numerical difference is not significant (t=0.348, df=354, p=0.7).

\subsection{Discussions}

A mismatch between the synthesized training data and the real recorded evaluation sets is the issue that needs tackling to improve classical MSS. What is required are improvements in both training and evaluation datasets.

Trying to get large training datasets with isolated instrument recordings for ensemble classical music is problematic. One possibility is not to attempt this, but to get isolated recordings without worrying about how this fits into an ensemble piece. It has been shown that for MSS of Western popular music, randomizing stems to create more diversity during training is useful \cite{jeon2024does}. This is even the case where the resulting mixture is unmusical. This approach could be tested for classical music.

Even so, more ensemble recordings are probably needed to represent passages in which musicians interact and blend their playing. To get good performances from classical musicians, they need to perform as an ensemble in the same space. That then makes it impossible to fully isolate instruments with traditional microphone techniques to get the ground truth. One solution might be to use DNNs to remove bleed from the microphone signals. 

When recording additional data, we should remember that classical music exists in a space with room acoustics and is recorded using many different microphone techniques. Additional data needs to support higher-order ambisonic formats to reflect the spatial information available in real-world listening environments and to allow different microphone configurations to be simulated. Multiple versions of the same piece under different reverberation conditions are also needed to enable generalisation to be tested across hall characteristics.

Another approach is to improve synthesized training data so that it is closer to real recordings. Professional music production tools include virtual instruments and convolution reverberators based on measurements of real instruments and spaces. These are very high quality, with increasing amounts of music for TV, film, podcasts and radio using these. The problem with current synthesized datasets (synthSOD, CadenzaWoodwind, EnsembleSet) is that they have used a limited number of virtual instruments and room reverbs, so there is insufficient diversity in the training set. The datasets also need to be better balanced between instruments to reflect real-world variability. The problem at the moment is there needs to be human intervention to get good quality audio. As professional music production tools move towards generative AI, generating additional diversity should become more straightforward.

Finally, improved processes for data augmentation show promise \cite{sarkar2023leveraging} to make the most of the limited data that is available. In this study, we only used two augmentation techniques, but pitch shifting and specAugmentation could also be very beneficial, for example. 
\section{Conclusion}

People with hearing loss could benefit from MSS of classical music, by enabling a personalised remix of the audio to improve the listening experience. But currently, MSS of classical music for real instruments, in ensembles larger than duos is unsolved. We used two synthesized music datasets to train eight MSS models, which between them targeted strings and woodwind instruments from ensembles of 2 to 5 instruments. A new dataset of synthesized woodwinds was developed for this work. We assessed the MSS performance on real recordings. Novelty compared to previous classical MSS comes from differences in the training data, the ConvTasNet MSS model used, and the exploration of causal and non-causal systems. 

Good performance was obtained for the synthesized validation set with an average SDR across instruments of 6.2 and 6.9 dB for the causal and non-causal systems. The separation achieved varied greatly across instruments. When applying the models on real data, however, separation was poor, dropping to 0.2 - 0.4 dB. This drop in performance is in line with what has been found in another recent paper. This generalisation issue could be due to several factors including the lack of instrument and reverberation diversity in the synthesized training data. It is well known that more training data can improve generalisation. However, before continuing to generate additional synthetic training data, a more robust protocol must be developed for gathering the audio to addresses current deficiencies. Furthermore, there is a lack of datasets of recorded, isolated instruments suitable for supervised learning or evaluation. Existing datasets are small, with insufficient variability in the instrumentation. Furthermore, the recording setups do not accurately reflect real performances where musicians play together, and this harms the quality of the music that results. Again, more data and improved benchmarks are needed.


\clearpage
\bibliographystyle{IEEEtran}
\bibliography{references}

\begin{thebibliography}{10}
\providecommand{\url}[1]{#1}
\csname url@samestyle\endcsname
\providecommand{\newblock}{\relax}
\providecommand{\bibinfo}[2]{#2}
\providecommand{\BIBentrySTDinterwordspacing}{\spaceskip=0pt\relax}
\providecommand{\BIBentryALTinterwordstretchfactor}{4}
\providecommand{\BIBentryALTinterwordspacing}{\spaceskip=\fontdimen2\font plus
\BIBentryALTinterwordstretchfactor\fontdimen3\font minus \fontdimen4\font\relax}
\providecommand{\BIBforeignlanguage}[2]{{%
\expandafter\ifx\csname l@#1\endcsname\relax
\typeout{** WARNING: IEEEtran.bst: No hyphenation pattern has been}%
\typeout{** loaded for the language `#1'. Using the pattern for}%
\typeout{** the default language instead.}%
\else
\language=\csname l@#1\endcsname
\fi
#2}}
\providecommand{\BIBdecl}{\relax}
\BIBdecl

\bibitem{hake2023development}
R.~Hake, M.~B{\"u}rgel, N.~K. Nguyen, A.~Greasley, D.~M{\"u}llensiefen, and K.~Siedenburg, ``Development of an adaptive test of musical scene analysis abilities for normal-hearing and hearing-impaired listeners,'' \emph{Behavior Research Methods}, vol.~15, no.~11, pp. 1--26, 2023, doi: 10.3758/s13428-023-02279-y.

\bibitem{moore2016effects}
B.~C. Moore, ``Effects of sound-induced hearing loss and hearing aids on the perception of music,'' \emph{Journal of the Audio Engineering Society}, vol.~64, no.~3, pp. 112--123, 2016, doi: 10.17743/jaes.2015.0081.

\bibitem{siedenburg2020can}
K.~Siedenburg, S.~R{\"o}ttges, K.~C. Wagener, and V.~Hohmann, ``Can you hear out the melody? testing musical scene perception in young normal-hearing and older hearing-impaired listeners,'' \emph{Trends in Hearing}, vol.~24, 2020, doi: 10.1177/2331216520945826.

\bibitem{Greasley2020}
A.~Greasley, H.~Crook, and R.~Fulford, ``Music listening and hearing aids: perspectives from audiologists and their patients,'' \emph{International Journal of Audiology}, vol.~59, no.~9, pp. 694--706, 2020, doi: 10.1080/14992027.2020.1762126.

\bibitem{roadabike2024}
G.~Roa-Dabike, M.~A. Akeroyd, S.~Bannister, J.~P. Barker, T.~J. Cox, B.~Fazenda, J.~Firth, S.~Graetzer, A.~Greasley, R.~R. Vos, and W.~M. Whitmer, ``{The first Cadenza challenges: using machine learning competitions to improve music for listeners with a hearing loss},'' \emph{IEEE Open Journal of Signal Processing}, 2025, to be published.

\bibitem{sisec2018}
F.-R. St{\"o}ter, A.~Liutkus, and N.~Ito, ``The 2018 signal separation evaluation campaign,'' in \emph{Latent Variable Analysis and Signal Separation}, 2018, doi: 10.1007/978-3-319-93764-9\_28.

\bibitem{vaisberg2019qualitative}
J.~M. Vaisberg, A.~T. Martindale, P.~Folkeard, and C.~Benedict, ``A qualitative study of the effects of hearing loss and hearing aid use on music perception in performing musicians,'' \emph{Journal of the American Academy of Audiology}, vol.~30, no.~10, pp. 856--870, 2019, doi: 10.3766/jaaa.17019.

\bibitem{Fabbro-2024}
G.~Fabbro, S.~Uhlich, C.-H. Lai, W.~Choi, M.~Martínez-Ramírez, W.~Liao, I.~Gadelha, G.~Ramos, E.~Hsu, H.~Rodrigues, F.-R. Stöter, A.~Défossez, Y.~Luo, J.~Yu, D.~Chakraborty, S.~Mohanty, R.~Solovyev, A.~Stempkovskiy, T.~Habruseva, N.~Goswami, T.~Harada, M.~Kim, J.~H. Lee, Y.~Dong, X.~Zhang, J.~Liu, and Y.~Mitsufuji, ``The sound demixing challenge 2023 – music demixing track,'' \emph{Transactions of the International Society for Music Information Retrieval}, Apr 2024, doi: 10.5334/tismir.171.

\bibitem{10446651}
W.~Tong, J.~Zhu, J.~Chen, S.~Kang, T.~Jiang, Y.~Li, Z.~Wu, and H.~Meng, ``Scnet: Sparse compression network for music source separation,'' in \emph{ICASSP 2024 - 2024 IEEE International Conference on Acoustics, Speech and Signal Processing (ICASSP)}, 2024, pp. 1276--1280, doi: 10.1109/ICASSP48485.2024.10446651.

\bibitem{10446843}
W.-T. Lu, J.-C. Wang, Q.~Kong, and Y.-N. Hung, ``Music source separation with band-split rope transformer,'' in \emph{ICASSP 2024 - 2024 IEEE International Conference on Acoustics, Speech and Signal Processing (ICASSP)}, 2024, pp. 481--485, doi: 10.1109/ICASSP48485.2024.10446843.

\bibitem{10096956}
S.~Rouard, F.~Massa, and A.~Défossez, ``Hybrid transformers for music source separation,'' in \emph{ICASSP 2023 - 2023 IEEE International Conference on Acoustics, Speech and Signal Processing (ICASSP)}, 2023, pp. 1--5, doi: 10.1109/ICASSP49357.2023.10096956.

\bibitem{musdb18-hq}
Z.~Rafii, A.~Liutkus, F.-R. Stöter, S.~I. Mimilakis, and R.~Bittner, ``Musdb18-hq - an uncompressed version of musdb18,'' doi: 10.5281/zenodo.3338373.

\bibitem{bonneville2013music}
A.~Bonneville-Roussy, P.~J. Rentfrow, M.~K. Xu, and J.~Potter, ``Music through the ages: Trends in musical engagement and preferences from adolescence through middle adulthood.'' \emph{Journal of personality and social psychology}, vol. 105, no.~4, p. 703, 2013, doi: 10.1037/a0033770.

\bibitem{9287146}
C.-Y. Chiu, W.-Y. Hsiao, Y.-C. Yeh, Y.-H. Yang, and A.~W.-Y. Su, ``Mixing-specific data augmentation techniques for improved blind violin/piano source separation,'' in \emph{2020 IEEE 22nd International Workshop on Multimedia Signal Processing (MMSP)}, 2020, pp. 1--6, doi: 10.1109/MMSP48831.2020.9287146.

\bibitem{stoter19}
F.-R. St\"oter, S.~Uhlich, A.~Liutkus, and Y.~Mitsufuji, ``Open-unmix - a reference implementation for music source separation,'' \emph{Journal of Open Source Software}, 2019, doi: 10.21105/joss.01667.

\bibitem{sarkar2022ensembleset}
S.~Sarkar, E.~Benetos, M.~Sandler \emph{et~al.}, ``Ensembleset: {A} new high-quality synthesised dataset for chamber ensemble separation,'' in \emph{International Society for Music Information Retrieval (ISMIR)}.\hskip 1em plus 0.5em minus 0.4em\relax Bengaluru, India: ISMIR, 2022, doi: 10.5281/zenodo.7316739.

\bibitem{li2018creating}
B.~Li, X.~Liu, K.~Dinesh, Z.~Duan, and G.~Sharma, ``Creating a multitrack classical music performance dataset for multimodal music analysis: {Challenges}, insights, and applications,'' \emph{IEEE Transactions on Multimedia}, vol.~21, no.~2, pp. 522--535, 2018, doi: 10.1109/TMM.2018.2856090.

\bibitem{10839019}
J.~Garcia-Martinez, D.~Diaz-Guerra, A.~Politis, T.~Virtanen, J.~J. Carabias-Orti, and P.~Vera-Candeas, ``Synthsod: Developing an heterogeneous dataset for orchestra music source separation,'' \emph{IEEE Open Journal of Signal Processing}, vol.~6, pp. 129--137, 2025.

\bibitem{CadenzaWoodwind}
G.~{Roa-Dabike}, T.~J. Cox, A.~J. Miller, B.~M. Fazenda, S.~Graetzer, R.~R. Vos, M.~A. Akeroyd, J.~Firth, W.~M. Whitmer, S.~Bannister, A.~Greasley, and J.~P. Barker, ``{The Cadenza Woodwind dataset: Synthesised quartets for music information retrieval and machine learning},'' \emph{Data in Brief}, vol.~57, p. 111199, 2024, doi: 10.1016/j.dib.2024.111199.

\bibitem{gotham2023openscore}
M.~Gotham, M.~Redbond, B.~Bower, and P.~Jonas, ``The ``openscore string quartet'' corpus,'' in \emph{Proceedings of the 10th International Conference on Digital Libraries for Musicology}.\hskip 1em plus 0.5em minus 0.4em\relax Milan, Italy: Association for Computing Machinery (ACM), 2023, pp. 49--57, doi: 10.1145/3625135.3625155.

\bibitem{jeon2024does}
C.-B. Jeon, G.~Wichern, F.~G. Germain, and J.~Le~Roux, ``Why does music source separation benefit from cacophony?'' in \emph{2024 IEEE International Conference on Acoustics, Speech, and Signal Processing Workshops (ICASSPW)}.\hskip 1em plus 0.5em minus 0.4em\relax IEEE, 2024, pp. 873--877, doi: 10.1109/ICASSPW62465.2024.10669899.

\bibitem{duan2011soundprism}
Z.~Duan and B.~Pardo, ``{Soundprism}: {An} online system for score-informed source separation of music audio,'' \emph{IEEE Journal of Selected Topics in Signal Processing}, vol.~5, no.~6, pp. 1205--1215, 2011, doi: 10.1109/JSTSP.2011.2159701.

\bibitem{shelley2010openair}
S.~Shelley and D.~T. Murphy, ``Openair: An interactive auralization web resource and database,'' in \emph{129th Audio Engineering Society Convention 2010}, 2010, pp. 1270--1278.

\bibitem{zotter2019ambisonics}
F.~Zotter and M.~Frank, \emph{Ambisonics: A practical 3D audio theory for recording, studio production, sound reinforcement, and virtual reality}.\hskip 1em plus 0.5em minus 0.4em\relax Springer Nature, 2019.

\bibitem{convtasnet}
Y.~Luo and N.~Mesgarani, ``Conv-tasnet: Surpassing ideal time–frequency magnitude masking for speech separation,'' \emph{IEEE/ACM Transactions on Audio, Speech, and Language Processing}, vol.~27, no.~8, pp. 1256--1266, 2019, doi: 10.1109/TASLP.2019.2915167.

\bibitem{sawata2021all}
R.~Sawata, S.~Uhlich, S.~Takahashi, and Y.~Mitsufuji, ``All for one and one for all: Improving music separation by bridging networks,'' in \emph{ICASSP 2021-2021 IEEE International Conference on Acoustics, Speech and Signal Processing (ICASSP)}.\hskip 1em plus 0.5em minus 0.4em\relax IEEE, 2021, pp. 51--55, doi: 10.1109/ICASSP39728.2021.9414044.

\bibitem{sarkar2023leveraging}
S.~Sarkar, L.~Thorpe, E.~Benetos, and M.~Sandler, ``Leveraging synthetic data for improving chamber ensemble separation,'' in \emph{2023 IEEE Workshop on Applications of Signal Processing to Audio and Acoustics (WASPAA)}.\hskip 1em plus 0.5em minus 0.4em\relax IEEE, 2023, pp. 1--5, doi: 10.1109/WASPAA58266.2023.10248118.

\end{thebibliography}

\end{document}